\documentclass[12pt,times]{elsarticle}

\usepackage[leqno]{amsmath}
\usepackage{amsthm}
\usepackage{graphicx}
\usepackage{epstopdf}
\usepackage{algorithm}
\usepackage{algorithmic}
\usepackage{subcaption,caption}
\usepackage{float} 
\usepackage{multirow}
\usepackage{lineno,color}
\usepackage[colorlinks=true, 
			linkcolor=cyan,
			citecolor=cyan, 
			urlcolor=cyan]
			{hyperref}
\numberwithin{equation}{section}

\newtheorem{theorem}{Theorem}[section]
\newtheorem{definition}{Definition}[section]    
\newtheorem{lemma}{Lemma}[section]

\newcommand{\ba}{\begin{array}}
\newcommand{\ea}{\end{array}}
\newcommand{\be}{\begin{equation}}
\newcommand{\ee}{\end{equation}}
 
\graphicspath{ {./Figures/} }

\def\R{\R}

\definecolor{blue}{rgb}{0,0,.65}
\definecolor{myblue}{rgb}{0,0,.5}
\definecolor{mygreen}{rgb}{0,.5,0}
\definecolor{red}{rgb}{1,0,0}
\definecolor{myred}{rgb}{.5,0,0}

\def\Rn{\mathbb{R}^n}
\def\Rm{\mathbb{R}^m}
\def\R{\mathbb{R}}
\def\Cmn{\mathbb{C}^{m\times n}}
\def\C{\mathbb{C}}
\def\Cn{\mathbb{C}^n}

\def\S{\mathbb{S}}
\def\cI{{\cal I}}

\def\a{{\bf a}}
\def\d{{\bf d}}
\def\p{{\bf p}}
\def\q{{\bf q}}
\def\t{{\bf t}}

\def\A{{\bf A}}

\def\T{{\bf T}}

\def\x{{\bf{x}}}
\def\y{{\bf{y}}}

 \date{}  
\begin{document}

\begin{frontmatter}




\title{Block Newton-type Algorithm for  Complex Compressed Sensing }

  \author[label1]{Hui Zhang}
  \ead{18118011@bjtu.edu.cn }
 \author[label1]{Xin Liu}
 \ead{19121573@bjtu.edu.cn}
 
 \author[label1]{Naihua Xiu}
 \ead{nhxiu@bjtu.edu.cn}
  \address[label1]{School of Mathematics and Statistics, Beijing Jiaotong University,
             Beijing, China}

\begin{abstract}
Many new challenges in Compressed Sensing (CS), such as block sparsity, arose recently. In this paper, we present a new algorithm for solving CS with block sparse constraints (BSC) in complex fields. Firstly, based on block sparsity characteristics, we propose a new model to deal with CS with BSC and analyze the properties of the functions involved in this model. We then present a new $\tau$-stationary point and analyze corresponding first-order sufficient and necessary conditions. That ensures we to further develop a block Newton hard-thresholding pursuit (BNHTP) algorithm for efficiently solving CS with BSC. Finally, preliminary numerical experiments demonstrate that the BNHTP algorithm has superior performance in terms of recovery accuracy and calculation time.
\end{abstract}
\end{frontmatter}
\section{Introduction}
\noindent Compressed Sensing (CS) was independently proposed by Cand\`es \citep{Cand2006a, Cand2006b} and D.Donoho\citep{Donoho2006}. It drives the development of a large number of interdisciplinary work, such as radar technology\citep{Karlina2011,Zhang2014,Xu2018}, medical imaging\citep{Lustig2007,Qaisar2013}, communication engineering technology\citep{Xie2013,Applebaum2010}, biosensor\citep{Sheikh2007}, spectral analysis \citep{Borgnat2008} and so on.  \\
\indent Given a measurement matrix, CS sample a sparse or compressible signal at a rate below the Nyquist theorem required.  Mathematically, the original CS model aim to recovering a vector $\x\in\Rn$ from measurement $\y\in\Rm$($m<n$) given by
\begin{eqnarray}
\min_{\x\in\Rn}& \|\x\|_0 \ \ s. t. \ \A\x=\y, \label{CS1}
\end{eqnarray}
where $\A$ is a measurement matrix of size $m\times n$ and $\ell_0$ ``norm'' of vector cuts its number of nonzero entries.  This model can also be treated as
\begin{eqnarray}
\min\limits_{\x\in\Rn}& \frac{1}{2}\|\A\x-\y\|^2 \ \ s. t. \ \|\x\|_0\leq s,\label{CS2}
\end{eqnarray}
where $s$ ($0<s\leq m$) reveal the sparsity. \\
\indent Due to the combinatorial nature of the $\ell_0$ norm, the above two problems involving the $\ell_0$ term are NP-hard\citep{Natarajan1995}. Some scholars have studied the continuous optimization model with convex or non-convex relaxation\citep{Tibshirani1996,Qaisar2013}.\\
\indent In recent years, sparse signal recovery problems often need to face at least two new challenges: (i) variables are complex\citep{Qaisar2013}. It is a core stem in the development of communication. Naturally, in the complex domain, models (\ref{CS1}) and (\ref{CS2}) have the following corresponding forms:
\begin{eqnarray}
\min\limits_{\x\in\Cn}& \|\x\|_0\ \ s. t. \  \A\x=\y, \nonumber\\
\min\limits_{\x\in\Cn}& \|\A\x-\y\|^2  \ \ s. t. \  \|\x\|_0\leq s,\nonumber
\end{eqnarray}
where $\A\in\Cmn $ is a complex sensing matrix.\\
\noindent (ii) variables are block sparsity. The classical sparsity model assumes that $\x$ has at most $s$ nonzero elements, however, it does not impose any further structure. As there are many cases in which the sparsity is subjected to blocks, meaning their sparsity comes in the form of regions, such as multi-band signals \citep{Mishali2009, Mishali2010, Landau1967, Mishali2011}, measurements of gene expression levels \citep{Parvaresh2008}, multiple measurement vector(MMV) problem \citep{Eldar2009a, Chen2006, Mishali2008, Eldar2009b}.  If we consider this kind of block sparsity, recovery may be possible in a more efficient way.\\
\indent To treat the two challenges mentioned above, in this paper, through a comprehensive study for block sparsity, we build a new CS model with block sparse constraints (BSC) in a complex field and develop a second-order algorithm for solving the new model. We summarize our main contributions as follows.\\
\indent {\bf I. New Model.} To better fit block sparsity prior information, we build a new CS model in a complex field. As far as we know, this is the first paper to explicitly takes block sparsity (also called sub-regional sparsity) into account in complex CS problem. By reducing the number of degrees of freedom of a complex sparse signal by permitting only certain configurations of the large and small coefficients, BSC models provide two immediate advantages to CS: a) during signal recovery, they enable us to better differentiate true signal information from recovery artefacts, which leads to a more robust recovery; b) they enable us to reduce, in some cases significantly, the number of measurements $A$ required to stably recover a signal. \\
\indent{\bf II. New theory.} We investigate and analyze the theoretical characteristics of CS with BSC. Despite the model being in a complex field, we present some nice propositions for the objective function, including continuity of gradients and descent lemma. Based on these, we propose a stationary point, which is called the $\tau$ stationary point. We then build first-order sufficient and necessary optimality conditions and an equivalent stationary equation. This is the first paper to investigate these theories in complex fields.\\
\indent{\bf III. New algorithm.} Based on the new theories mentioned above, we design a second-order method, called block Newton hard thresholding pursuit(BNHTP), for solving CS with BSC.  Although BNHTP is a second-order method, it still maintains an efficient operating speed as solving Newton equation in subspace. And hence it can deduce the calculation time. Benefiting from block sparsity, BNHTP enables us to implement in a distributed and parallel manner and is thus capable of solving large-scale distributed/decentralized problems. The BNHTP is suitable to treat the signal with sparsity subjected to blocks. Numerical shows its advancement than AMP dealt with the conventional sense, ignoring the additional structure in the problem.\\
\indent To carry out these works, we organize this paper as follows. In the next section,  we introduce some relevant works for CS. In Section 3,  we present our CS model with BSC and its useful propositions as well as optimality conditions. In Section 4, we propose some details for our BNHTP used to solve CS with BSC. In Section 5, we demonstrate the high performance of BNHTP by conducting extensive numerical comparisons among AMP for solving CS problems. Some conclusive remarks are given in the last section.\\
\section{Relavant works}
\noindent To the best of our knowledge, there is no study dedicated exclusively to CS with BSC in the complex field. However, an impressive body of work has developed various similar studies for CS. In this section, we conduct a brief review of relevant and popular works in CS.\\
\indent As for classical CS, scholars have made a series of core achievements in the direction of theory, such as RIP conditions\citep{Cand2006b}, coherence\citep{Kutyniok2013}, and other related sparse optimization theories\citep{Cotter2005, Zhou2019, Zhou2021}. Based on these results, many practical and effective classical algorithms have been proposed continuously, such as basic pursuit (BP) algorithm\citep{Chen2001}, matching pursuit (MP) algorithm\citep{Mallat1993}, approximate message pass (AMP) algorithm\citep{Donoho2009} and other first-order algorithms. These first-order algorithms have the advantages of simple structure, small storage and easy calculation. However, these algorithms have the drawbacks of slow convergence and low accuracy. Fast and high-accuracy algorithms have become the research direction. Some progress and achievements have been made, such as EM algorithm\citep{Krishnapuram2005}, QWL-QN algorithm\citep{Andrew2007}, SSNAL algorithm\citep{Li2016}, NL0R algorithm\citep{Zhou2021}, NHTP algorithm\citep{Zhou2019} and other second-order algorithms. Despite the emergence of many algorithms, the AMP algorithm is still the most used in the industry, and it has also become an algorithm to determine whether a new model or algorithm is good.\\ 
\indent As for the sparsity, Mishali and Eldar first introduced group sparsity into CS\citep{Eldar2009a}. The main idea of group sparsity is to divide the signal vector into several groups, and then transfer the classical non-zero elements as few as possible to the non-zero groups as few as possible. Since then, many methods were proposed to solve this type of CS and achieved a good reconstruction performance. There is three main steam in this direction. The first body is a convex method. In \citep{Eldar2009a}, as an extension of BP method, the mixed $l_1$ and $l_2$ norm algorithm is proposed. Subsequently,  many iterative algorithms have been carried out for recovering block sparse signals \citep{Zeinalkhani2015}, which solves a weighted  $l_1$ mixed $l_2$ minimization in each iteration. The above two methods have good quality, but they are slow. Another convex method called BSL0 \citep{Hamidi2010} is also tested to be effective. The second family of approaches are greedy algorithms, such as block OMP (BOMP) \citep{Eldar2010}, block CoSaMP \citep{Baraniuk2010}, and block StOMP \citep{Huang2015}, they are very fast, but the sparsity should be known as a prior. The third kind of approach is non-convex methods \citep{Wang2013, Yin2013}, it is shown that the non-convex methods surpass the mixed $l_1$ and $l_2$ norm algorithm. However, all these works treat each group equally. As far as we know, there are no studies on exploiting the sparsity of each region. To fill this gap, a new model will be proposed in this paper.

\section{Model and optimality theories}
This section mainly focuses on proposing and analysing a new CS model. The relevant proofs of results are provided in Appendix.\\
\indent At first, apart from the aforementioned notation, we also summarize some other symbols here.
Throughout the paper, we use ``$:=$'' to mean ``define''. For an index set $\T$, let $|\T|$ be the number of elements of $\T$ and $ \T_{C}:=\{1,2,\dots,\}\setminus \T$ be its complementary set. For matrix $A$, we write $a_i$, $A_i$, and $A_{ij}$ as the  the ith row, the jth column, and $(i, j)$th entry of $A$. Let $\|\bullet\|$ denote the Frobenius norm for matrices and the Euclidean norm for vectors. The identity matrix is denoted
by I.\\
\indent We treat $x$ as consisting of blocks with given lengths $d_i$, $1\leq i\leq I$. Denoting by $x_{[i]}$ the $i$-th sub-block of length $d_i$, we can write $x$ as
\[x = [
\underbrace{x_1,x_2,...,x_{d_1}}_{x_{[1]}},\underbrace{x_{d_1+1},...,x_{d_1+d_2}}_{x_{[2]}},...
,\underbrace{x_{n-d_I}...,x_{n}}_{x_{[I]}}]^{\top}.\]
We call $x$ a block sparse if $\|\x_{[i]}\|_0\leq s_{i}$ holds for any $1\leq i\leq I$. When $I=1$, block sparsity reduces to conventional sparsity. For easy to express, denote $\cI:=\{1,2,...,I\}$ and $\S$ be the block sparse set: \[\S:=\{\x\in\Cn \mid \x_{[i]}\in \S_{[i]}, i = 1,2, ..., I\},\]
where $\S_{[i]}:=\{\t\in \C^{d_{i}} \mid \|\t\|_0\leq s_{i}\}$ and $s_{i}$ is a positive integer.\\
\indent Similarly, we can write $\A\in\Cmn$ as a concatenation of column-blocks $\A_{[i]}$ of size $m\times d_i$:
\[\A =[\underbrace{\a_1,\a_2,...,\a_{d_1}}_{\A_{[1]}},\underbrace{\a_{d_1+1},...,\a_{d_1+d_2}}_{\A_{[2]}},...
,\underbrace{\a_{n-d_I}...,\a_{n}}_{\A_{[I]}}].\]
Hence, it is easy to see that $\A\x = \sum\limits_{i=1}^{\boldsymbol{I}}\A_{[i]} \x_{[i]}$.\\
Based on the above notions, we build a special CS with block sparse constraint 
\be\label{BCS}
\min\limits_{\x\in\Cn}  f(\x), ~ s.t.~ ~x\in \S,
\ee
where $f(\x):=\sum\limits_{i=1}^{\boldsymbol{I}}\|\A_{[i]} \x_{[i]}-\y\|^{2}$, $i\in \cI$.\\
\indent We will focus on the model (\ref{BCS}) in this paper. At first, we investigate the propositions of the complex function $f(\x)$ as the following three lemmas.
\begin{lemma}\label{f-Lips}
For the $f(\x)$ in model (\ref{BCS}), there exist a constant $\alpha_{f}=\dfrac{1}{2 \lambda_{\max }(\A^{\rm{H}} \A)}$, such that for $\forall \p, \q \in \Cn$,
\[\left\|\nabla f(\p)-\nabla f(\q)\right\| \leq \frac{1}{\alpha_{f}}\left\|\p-\q\right\|.\]
\end{lemma}
\noindent {\bf Remark.} Lemma \ref{f-Lips} indicates the gradient of the objective function $\nabla f(\x)$ is Lipschitz continuous with constant $2 \lambda_{\max }(\A^{\rm{H}} \A)$, which is a core proposition to analyse optimality for model (\ref{BCS}).
\begin{lemma}\label{f-descent}
For $f(\x)$ in model (\ref{BCS}), $\forall \p, \q \in \Cn$, there exists a constant $\alpha_{f}$ such that
\[f(\p) \leq f(\q)+(\p-\q)^{\rm{H}} \nabla f(\q)+(\overline{\p}-\overline{\q})^{\rm{H}} \overline{\nabla f(\q)}+\frac{1}{\tau}\left\|\p-\q\right\|^{2}, \forall \tau \leq \alpha_{f}/2.\]
\end{lemma}
To present optimization condition of model (\ref{BCS}), we further need a convexity (in complex) of $f(\x)$.
\begin{lemma}\label{lemma3}
For the $f(\x)$ in model (\ref{BCS}), we have
\begin{align*}
f(\p)\geq f(\q)+ (\p-\q)^{\rm{H}} \nabla f(\q)+  (\overline{\p}-\overline{\q})^{\rm{H}} \overline{\nabla f(\q)},\quad \forall \p, \q \in \Cn.
\end{align*}
\end{lemma}
\indent To proceed, we need to present a stationary point for the model (\ref{BCS}).
\begin{definition}\label{lsc}
Let $\x^{*} \in \Cn$ be a feasible point of model (\ref{BCS}). We say that $\x^{*}$ is a $\tau$-stationary point if there exists a $\tau > 0$, such that
\begin{eqnarray}
\x^{*}_{[i]}\in{\rm P}_{{S}_{[i]}}((\x^{*}-\tau \nabla f (\x^{*}))_{[i]}), \forall i\in \cI.
\end{eqnarray}
\end{definition}
\indent Now, by the above results, we are ready to give the first-order necessary and sufficient condition for the model (\ref{BCS}).
\begin{theorem}\label{first-order}
Let $\x^{*}\in\Cn$  be a feasible point of model (\ref{BCS}). The $\x^{*}$ is the optimal solution of the model (\ref{BCS}) if and only if it is a $\tau$-stationary point.
\end{theorem}
This theorem ensures we solve the model (\ref{BCS}) by solving the s-stationary point. However, the s-stationary point is still not very easy to use. We hence need to present an equivalent form of $\tau$-stationary point. The following lemma shows this relationship.
\begin{lemma}\label{Stationary-or}
Let $\x^{*} \in \Cn$ be a feasible point of model (\ref{BCS}). The $\x^{*}$  is a $\tau$-stationary point if and only if\\
(i) If ${\|\x_{[i]}^{*}}\|_{0} = s_{i}$, then
\begin{align*}
 \left|\nabla f(x^*)_{j}\right| \begin{cases}=0, & j \in \Gamma_{[i]}, \\ \leq M_{s_i}(\left|\x^{*}_{[i]}\right|) / \tau, & j \notin \Gamma_{[i]}.\end{cases}
\end{align*}
(ii) If ${\|\x_{[i]}^{*}}\|_{0} < s_{i}$, then
\begin{align*}
\nabla f(x^*) _{[i]}=0,\quad  M_{s_i}(|\x^{*}_{[i]}|)=0 .
\end{align*}
where $\Gamma_{[i]}$ is supporting index set of non-zero element of $\x_{[i]}^{*}$, and $M_{s_i}(\t)$ represents the $s_i$ the maximum element of $|\t|$.
\end{lemma}
\indent To end this section, we simply the expression of the stationary equation introduced in lemma \ref{Stationary-or}. Defined the whole support set as follows:
\begin{eqnarray}
{\cal T}(\x ; \tau):=\bigcup \T_{[i]} ,\label{ChooseT}
\end{eqnarray}
where
\[\T_{[i]} = \left\{\T'\subset\{d_i+1,...,d_i+d_{i+1}\} \mid |\T'|=s_i, |\bar{\x}_{j}| \geq M_{s_i}(|\bar{\x}_{[i]}|), \forall j \in \T'\right\}\]
and
$\bar{\x}:=\x-\tau \A^{\rm{H}}(\A \x-\y)$.\\
Hence, we can easily get if $\x$ is an s-stationary point if and only if
\begin{eqnarray}
F_{\tau}(\x ; \T):=\left[\begin{array}{c}
\nabla_\T f(\x) \\
\x_{\T_{C}}
\end{array}\right]=0, \forall \T\in{\cal T}. \label{stap}
\end{eqnarray}

\section{Algorithm} \label{Section-algorithm}
This part mainly proposes an algorithm to solve the model (\ref{BCS}).\\
\indent Based on (\ref{stap}) corresponding to the $\tau$-stationary point, the BNHTP algorithm can be divided into three main steps: (a) choose a suitable support set, then (b) select an appropriate search direction, and (c) generate a new iteration point by a line search. We summarise the specific iterative steps of the BNHTP in Algorithm \ref{BN-algo}.
 \begin{algorithm}[!th]
	\caption{BNHTP for CS  with BSC \eqref{BCS}}
	\begin{algorithmic}[1]
	\STATE \textbf{Require} sensing matrix $\A\in \C^{m \times n}$, observation vector $\y\in \C^{m}$, and maximum number of iteration $K$.
	\FOR{$k=0,1,2,\cdots,K$}
	\STATE Calculate support set $\T^k\in{\cal T}(\x^k ; \tau)$ defined in (\ref{ChooseT}).
	\STATE Calculate descent direction $\d^{k}$ by (\ref{BCS5}).
	 \STATE Update $\x^{k+1}=\left[\begin{array}{c}\x^{k}_{\T^{k}}+\ \alpha^{k}\d^{k}_{\T^k} \\ 0\end{array}\right]$, where $\alpha^{k}$ is obtained through $\operatorname{Armijo}$ line search.
	\STATE \textbf{if} $\operatorname{Tol}_{\tau}(\x^{k} ; \T)<\varepsilon$,
	\textbf{then} stop
\ENDFOR
\RETURN $\x^{k+1}$
	\end{algorithmic}\label{BN-algo}
\end{algorithm}

The following remarks are the details of each step in the BNHTP algorithm.\\
\noindent {\bf Remark (i)} For the halting condition of BNHTP in Algorithm 1, in our numerical experiments, we set $\operatorname{Tol}_{\tau}(\x^{k}; \T^k)$ by
\[\operatorname{Tol}_{\tau}(\x^{k}; \T^k):=\left\|F_{\tau}(\x^{k} ; \T^k)\right\|+\max _{i \in \T^k_{C}}\left\{\max (\left|\nabla_{i} f(\x^{k})\right|-\frac{M_{l}(\left|\x^{k}\right|)}{\tau}, 0)\right\},\]
where $M_{l}(\left|\x\right|)$ represents the largest element in $\boldsymbol {x}$, $l$ is the number of indicators in the support set.\\

\noindent {\bf Remark (ii)} In step 4, we aim to find a suitable direction to solve \eqref{stap}. There are two methods to carry out. \\
a) By using $-F_{\tau}(\x ; \T^k)$ and the idea of momentum, we can find a directly descent direction $\d_g^k$ i.e.,
\[\d_g^k:=\left[\begin{array}{c}
\A_{\T^k}^{\rm{H}}(\A_{\T^k} \x_{\T^k}^k-\y)+\eta \d^{k-1}\\
-\x_{{\T}^{k}_{C}}^{k}
\end{array}\right],\]
where $\eta>0$.\\
b) By the idea of the Newton method, we can also get the Newton search direction $\d_{{N}}^{k}$ from the equation system (\ref{stap}). The Newton equation can be calculated as
\[
\nabla F_{\tau}(\x^{k} ; \T^{k})\d_{\rm{N}}^{k}=-F_{\tau}(\x^{k} ; \T^{k}),\]
where
\[
\nabla F_{\tau}(\x^{k} ; \T^{k})=\left[\begin{array}{cc}
\nabla_{\T^{k}, \T^{k}}^{2} f(\x^{k}) & \nabla_{\T^{k}, \T_{C}^{k}}^{2} f(\x^{k}) \\
0 & I
\end{array}\right].\]
Then we have
$$
\left[\begin{array}{cc}
\nabla_{\T^{k}, \T^{k}}^{2} f(\x^{k}) & \nabla_{\T^{k}, \T_{C}^{k}}^{2} f(\x^{k}) \\
0 & I
\end{array}\right]\left[\begin{array}{l}
(\d_{N}^{k})_{\T ^k} \\
(\d_{N}^{k})_{\T_{C}^{k}}
\end{array}\right]=-\left[\begin{array}{c}
\nabla_{\T^{k}} f(\x^{k}) \\
\x_{\T_{C}^{k}}^{k}
\end{array}\right]
 $$
By simple tidying up, we can get
\[
\d_{N}^{k}=\left[\begin{array}{c}
((\A^{\rm{H}} \A)_{\T^{k},\T^{k}})^{-1}((\A^{\rm{H}} \A)_{\T^{k}, \T^{k}_{C}} \x_{\T^{k}_{C}}^{k}-(\A^{\rm{H}}(\A \x^{k}-\y))_{\T^{k}}) \\
-\x_{\T_{C}^{k}}^{k}
\end{array}\right].
\]
From the perspective of optimization, in the initial iteration of the algorithm, if the initial point is far from the optimal solution, the direct use of $\d_N^k$ is often not as effective as selecting the $\d_g^k$. Setting Newton's direction $\d_N^k$ can achieve the optimal solution faster and better when it is close to the optimal solution. From the experience of the algorithm in the actual number field, if there is only a Newton direction, the global convergence of the algorithm can not be guaranteed. Therefore, in BNHTP, the Newton steps are only performed when one of the following conditions is satisfied,
\begin{eqnarray}
\left\langle\nabla f(\x^{k})_{\T^{k}}, (\d_{N}^k)_{\T^{k}}\right\rangle \leq-\gamma\left\|\d_{N}^k\right\|^{2}+\left\|\x_{{\T}^{k}_{C}}^{k}\right\|^{2} /(4 \tau),\label{BCS66}
\end{eqnarray}
where $\gamma, \tau\in(0,1)$. The above condition can be deemed as the conditions for checking the neighbourhood of a locally (or globally) optimal solution. Once a point $\x^k$ meets the above conditions, it falls into the neighbourhood of a locally (or globally) optimal solution. Then we perform Newton's steps to find it quickly. In a nutshell, using \eqref{BCS66} as a switch of Newton steps enables to accelerate the convergence. To sum up, we use the search direction as
\begin{eqnarray}
\d^{k} =
\begin{cases} \d^{k}_{N}, & \mbox{if } (\ref{BCS66})~\mbox{is satisfied}, \\ \d^{k}_{g}+\eta\d^{k-1}, & \mbox{otherwise}.
\end{cases}\label{BCS5}
\end{eqnarray}

\noindent {\bf Remark (iii)}  In step 5, $\alpha^{k}=\beta^\ell$ is obtained by $\operatorname{Armijo}$ line search, where $ \ell$ is the smallest positive integer in $\{0,1,2,\cdots\}$, such that
$$
\begin{aligned}
f(\x^{k}+ \beta^{\ell} \d^{k})
\leq f(\x^{k})+\sigma \beta^{\ell}\left[(\d^{k})^{\rm{H}} \A^{\rm{H}}(\A \x^{k}-\y)+(\d^{k})^{\rm{T}} \A^{\rm{T}}\overline{\A \x^k}-\overline{\y})\right].
\end{aligned}
$$
The purpose of obtaining $\alpha^{k}$ by line search is to ensure that the objective function can be fully reduced.
\section{Numerical Experiments} \label{Section-Numerical}
In the previous chapter, a new BNHTP algorithm was designed to solve the model \eqref{BCS}. This part mainly compares the BNHTP algorithm with the AMP algorithm and shows the relevant experimental results in the complex field to verify the superiority of the BNHTP $\footnote{Our source code is available at https://github.com/BJTUZhangHui/BNHTP$\_$for$\_$CS.git}$. The code uses MATLAB (R2019a) on a laptop with 8GB memory and a 3.3GHz CPU. \\
\indent Numerical experiment mainly shows the efficiency and accuracy of two algorithms, in which the different effects of data type, sparsity $\bar s$ and noise. The numerical experiment also controls the probability of false alarms, the influence of changing the noise size on the data, observes the probability of missed detection, sets the size and change of the corresponding threshold, and further compares the stability of the algorithm.
\subsection{Data sets}
\indent We get the data sets from wireless communication. Set received signal
\[\y=\A\x+\boldsymbol{z},\]
where $\y \in \C^{839 \times 1}$, $\A \in \C^{839 \times 2048}$. The elements in $\x \in \C^{2048 \times 1}$ are divided into 64 groups, which means $I=64$ in our model. Each group corresponds to a communication user, and each group has 32 elements, of which the 1st to 32nd elements of $\x$ are the first group, the 33rd to 64th elements of $\x$ are the second group, corresponding to user 2, and so on until the 64th user. The user activates with a certain probability, Users are activated with a certain probability, and we denote the total number of activated users by $\bar s$, where $0\leq {\bar s}\leq64$. If a user activates, only one of the 32 elements corresponding to the user is non-zero, which means $s_i=1$ for $i\in \{1,2, ..., I\}$ in our model. The position of the non-zero term is randomly selected from the 32 elements. In contrast, the value of the non-zero term obeys the complex Gaussian distribution $\mathcal{C} \mathcal{N}(0, \beta)$. If a user is inactive, the 32 elements corresponding to that user are zero. The noise vector $\boldsymbol{z} \in \C^{839 \times 1}$ obeys complex Gaussian distribution $\mathcal{C N}(0, \sigma^{2} \rm{I})$.\\
\indent There are four generation methods to design the sensing matrix $\A$, including i)\A1: gaussian complex matrix, ii)\A2: partial discrete cosine transform complex matrix, iii)\A3: exponential type (I) complex matrix and iv)\A4: exponential type (II) complex matrix. The mathematical expressions and characteristics of these four sensing matrices are given in appendix E.\\
\indent To facilitate the comparison of numerical experiments, six evaluation indicators are given below.
(i) Iter: Number of iterations; (ii) Time(s): CPU time; (iii) R-error: Relative error, i.e., \[{\|\x^{\prime}-\x^{*}\|}/{\left\|\x^{\prime}\right\|},\]
where $\x^{*}$ is the true signal, $\x^{\prime}$ is the recovered signal; (iv) Obj-value: Objective value $f(\x^{*})$; (v) T rate: the correct recovery rate for the position of the non-zero element; (vi) Tc rate: the correct recovery rate for the position of the zero element.\\
\indent The relative error reflects the similarity between the recovered signal and the real signal. The larger the T recovery rate and the Tc recovery rate are, the better the position of the non-zero element and the zero elements of the recovered signal and the real signal. This is the best way to achieve the desired effect. Combined with the application background of data communication, T recovery rate and Tc recovery rate can also reflect the user's activation and inactivation states to a certain extent.
\subsection{Numerical results}
When $\bar s$= 20, $\sigma=0.001$, under the same four types of sensing matrices and real signals, two algorithms are executed 5 times and then averaged. The experimental results are as follows:
\begin{table}[H]
\centering
\caption{ Two Algorithm for four kinds of sensing Matrix}
\begin{tabular}{cccccccc}
\hline Matrix & Method &  Iter & Time(${s}$) & R-error &  Obj-value & T rate (\%) & Tc rate(\%) \\
\hline
\multirow{2}{*}{\A1}
&AMP   & 9.4   & 0.12    & 0.048   & 0.421  &  99.8 & 100\\
&BNHTP & {\bf 3.2}   & {\bf 0.04 }   & {\bf 0.035}   & {\bf 0.424}  & 99.8 & 100\\
\hline
\multirow{2}{*}{\A2}
&AMP   & 15.2   & 0.26    & 0.068   & 0.413  & 98  & 99.93 \\
&BNHTP & {\bf 3.3}    & {\bf 0.04}    &{\bf 0.032}   & {\bf 0.416}  & {\bf 100} & {\bf 100}\\
\hline
\multirow{2}{*}{\A3}
&AMP   & 10.1   & 0.11    & 0.051   & 0.424  & 99.8 & 100 \\
&BNHTP &{\bf  3.5}    &{\bf  0.07}    & {\bf 0.033}   &{\bf  0.413} & {\bf 100} & 100\\
\hline
\multirow{2}{*}{\A4}
&AMP   & 27    & 0.71    & 0.082   & 0.455  & 99.4 & 99.97 \\
&BNHTP & {\bf 3}   & {\bf 0.03}    & {\bf 0.035}   & {\bf 0.418}  & {\bf 100}  & {\bf 100}\\
\hline
\end{tabular}
\end{table}
\indent In the above numerical experimental results for the four types of sensing matrices, the number of iterations of the BNHTP algorithm is less than the result processed by the AMP algorithm, and the corresponding CPU time and relative error are also smaller. The recovery rate is better than, which also shows that the algorithm has obvious advantages in the recovery of signal non-zero elements and zero-element positions.

The above experiment fixed the sparsity $\bar s$=20 in the data set to test the performance of the algorithm. The following considers changing the sparsity s, that is, changing the number of non-zero elements of the signal. Also when $\sigma=0.001$, the same four types of experimental results are under the sensing matrix and the real signal. Among them, 100 experiments are performed for different $\bar s$ in the experiment, and the number of iterations, CPU time, relative error, target value, T recovery rate and Tc recovery rate are averaged values. The experimental results are as Tables \ref{diff-s1}-\ref{diff-s4}.
\begin{table}[H]
\centering
\caption{ Two Algorithms for Gaussian Matrix with different $s$}
\begin{tabular}{cccccccc}
\hline $\bar s$ & Method &  Iter & Time(${s}$) & R-error &  Obj-value & T rate (\%) & Tc rate(\%) \\
\hline
\multirow{2}{*}{10}
&AMP  & 9.23  & 0.102  & 0.0481 & 0.414  & 100.00 & 99.90 \\
&BNHTP &{\bf  3.00}  & {\bf 0.037} & {\bf 0.0304 }  & {\bf 0.414} & 100.00 & {\bf 100.00} \\
\hline
\multirow{2}{*}{20}
&AMP   &  9.20  & 0.098 & 0.0603   & 0.420 &95.00 & 99.90 \\
&BNHTP & {\bf 3.54}  & {\bf 0.042} & {\bf 0.0326} & {\bf 0.409} & {\bf 100.00} & {\bf 100.00} \\
\hline
\multirow{2}{*}{30}
&AMP   & 10.02 & 0.107 & 0.0563   & 0.429   & 96.97 & 99.90 \\
&BNHTP & {\bf 3.25}  &{\bf  0.042} & {\bf 0.0338} & {\bf 0.405} & {\bf 100.00} & {\bf 100.00} \\
\hline
\multirow{2}{*}{40}
&AMP   &  10.05 & 0.108  & 0.0561 & 0.425  & 99.95 & 99.88 \\
&BNHTP & {\bf 3.44}  & {\bf 0.041}& {\bf 0.0353}& {\bf 0.402} & {\bf 100.00} & {\bf 100.00} \\
\hline
\multirow{2}{*}{50}
&AMP   & 10.86 & 0.124  & 0.0538 & 0.433 & 99.40 & 99.88 \\
&BNHTP & {\bf 4.56}  & {\bf 0.057}& {\bf 0.0365} &{\bf  0.387} & {\bf 99.58 } & {\bf 99.91} \\
\hline
\multirow{2}{*}{60}
&AMP   & 11.38 & 0.128  & 0.0569 & 0.437 & 97.37 & 98.53\\
&BNHTP & {\bf 4.75}  & {\bf 0.059} & {\bf 0.0392} & {\bf 0.392} & {\bf 97.75}  & {\bf 99.93} \\
\hline
\end{tabular}\label{diff-s1}
\end{table}

\begin{table}[H]
\centering
\caption{ Two Algorithms for Partial Discrete Cosine Matrix with different $\bar s$}
\begin{tabular}{cccccccc}
\hline $\bar s$ & Method &  Iter & Time(${s}$) & R-error &  Obj-value & T rate (\%) & Tc rate(\%) \\
\hline
\multirow{2}{*}{10}
&AMP  & 19.60 & 0.199 & 0.0502 & 0.421   & 100.00 & 99.95 \\
&BNHTP & {\bf 3.04} & {\bf 0.032}   & {\bf 0.0463}   & {\bf 0.415}   & 100.00 & {\bf 100.00} \\
\hline
\multirow{2}{*}{20}
&AMP   &  10.57 & 0.119   & 0.053 & 0.424   & 100.00 & 99.95 \\
&BNHTP &  {\bf 3.01}  & {\bf 0.034}   & {\bf 0.030}   & {\bf 0.410}   & 100.00 & {\bf 100.00} \\
\hline
\multirow{2}{*}{30}
&AMP   & 12.97 & 0.138 & 0.0494 & 0.428   & 99.97 & 99.95 \\
&BNHTP & {\bf 5.84} & {\bf 0.072}   & {\bf 0.0402}   & {\bf 0.403}   & 98.07 & {\bf 99.97} \\
\hline
\multirow{2}{*}{40}
&AMP   &  12.49 & 0.138 & 0.0484 & 0.429   & 95.78 & 99.95 \\
&BNHTP & {\bf 5.17} & {\bf 0.068}   & {\bf 0.0359}   & {\bf 0.400}   & {\bf 99.50} & {\bf 99.99} \\
\hline
\multirow{2}{*}{50}
&AMP   & 18.99 & 0.196 & 0.0607 & 0.437   & 96.68 & 99.93 \\
&BNHTP & {\bf 5.46} & {\bf 0.076}   & {\bf 0.0344 }  & {\bf 0.395}   & {\bf 97.60} & {\bf  99.94} \\
\hline
\multirow{2}{*}{60}
&AMP   & 28.40 & 0.280 & 0.0698 & 0.448   & 97.77 & 99.88 \\
&BNHTP & {\bf 5.84} & {\bf 0.083}   & {\bf 0.0339}   & {\bf 0.388}  & {\bf 99.75} &{\bf  99.99} \\
\hline
\end{tabular}\label{diff-s2}
\end{table}
\indent We also test the BNHTP under different sparsity $\bar s$ in Tables \ref{diff-s1}-\ref{diff-s4}. The results processed by the BNHTP algorithm are significantly less than AMP in terms of the number of iterations. The BNHTP also has advantages in CPU time and relative error. As the degree $\bar s$ increases gradually, the T and Tc recovery rates are affected. The BNHTP algorithm is still better than the AMP algorithm on T recovery rate and Tc recovery.\\
\indent To better test the performance of the evaluation algorithm in processing such data, the false alarm probability (denoted as FAP) and the false identification rate ((denoted as FIR)) need to be introduced. In 10000 repetitive simulations, User 1 is inactive in each simulation (that is, all 32 elements are zero). Still, after detection by the algorithm, among the 32 elements corresponding to User 1 in the 100 simulation results, there is a non-zero in User 1. In other simulation results, the 32 elements corresponding to User 1 are all zero, so the false alarm probability of this user is 1\%. 
\begin{table}[H]
\centering
\caption{ Two Algorithms for exponential type (I) matrix with different $\bar s$}
\begin{tabular}{cccccccc}
\hline $\bar s$ & Method &  Iter & Time(${s}$) & R-error &  Obj-value & T rate (\%) & Tc rate(\%) \\
\hline
\multirow{2}{*}{10}
&AMP  & 9.97  & 0.104   & 0.0513   & 0.426   & 99.90 & 99.98\\
&BNHTP & {\bf 2.04} &   {\bf 0.022}   &  {\bf 0.0339}   & {\bf 0.414}   &  {\bf 100.00} &  {\bf 100.00} \\
\hline
\multirow{2}{*}{20}
&AMP   &  10.57 & 0.119   & 0.0531   & 0.424   & 100.00 & 99.95 \\
&BNHTP &  {\bf 3.29 }&  {\bf 0.039  } &  {\bf 0.0363}   &  {\bf 0.408}  & 99.90 &  {\bf 100.00} \\
\hline
\multirow{2}{*}{30}
&AMP   & 10.40 & 0.108   & 0.0566   & 0.438   & 100.00 & 99.98 \\
&BNHTP &  {\bf 3.00} &  {\bf 0.033 }  &  {\bf 0.0339}   &  {\bf 0.407}   & 100.00 &  {\bf 100.00 } \\
\hline
\multirow{2}{*}{40}
&AMP   &  10.88 & 0.115   & 0.0490   & 0.429   & 99.95 & 99.98 \\
&BNHTP &  {\bf 3.03} &  {\bf 0.036}   &  {\bf 0.0328}   &  {\bf 0.398}   &  {\bf 100.00} &  {\bf 100.00} \\
\hline
\multirow{2}{*}{50}
&AMP   & 11.03 & 0.146   & 0.0533   & 0.430   & 100.00 & 99.98 \\
&BNHTP &  {\bf 4.14} &  {\bf 0.055}   &  {\bf 0.0297}   &  {\bf 0.396}   & 99.98 &  {\bf 100.00} \\
\hline
\multirow{2}{*}{60}
&AMP   & 11.09 & 0.116   & 0.0594   & 0.437   & 99.87 & 99.98 \\
&BNHTP &  {\bf 6.96} &  {\bf 0.113}   &  {\bf 0.0321}   & {\bf  0.390}   & {\bf  99.63} &  {\bf 99.99} \\
\hline
\end{tabular}\label{diff-s3}
\end{table}

\begin{figure}
\centering
\includegraphics[height=7cm,width=7cm]{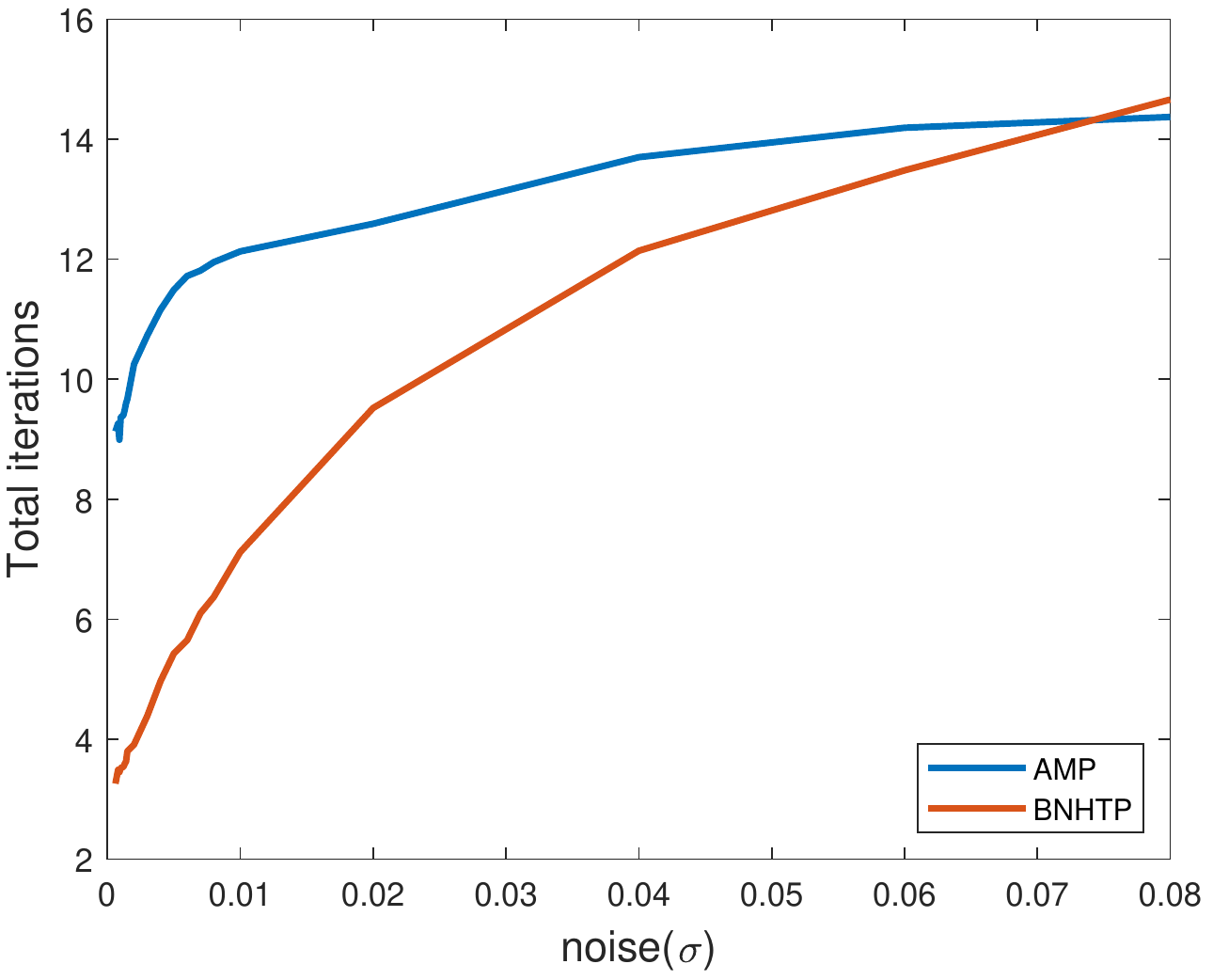}
\includegraphics[height=7cm,width=7cm]{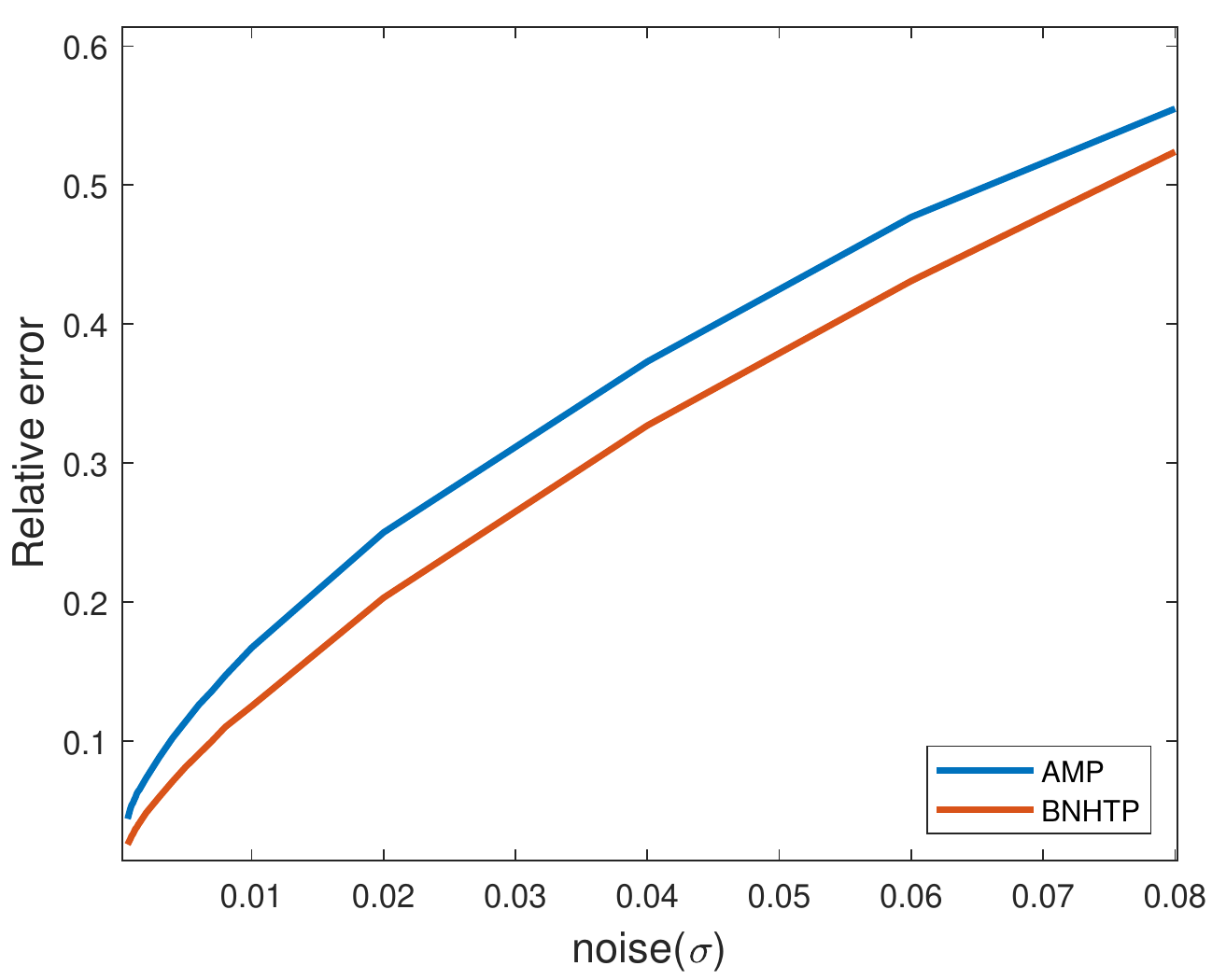}
\includegraphics[height=7cm,width=7cm]{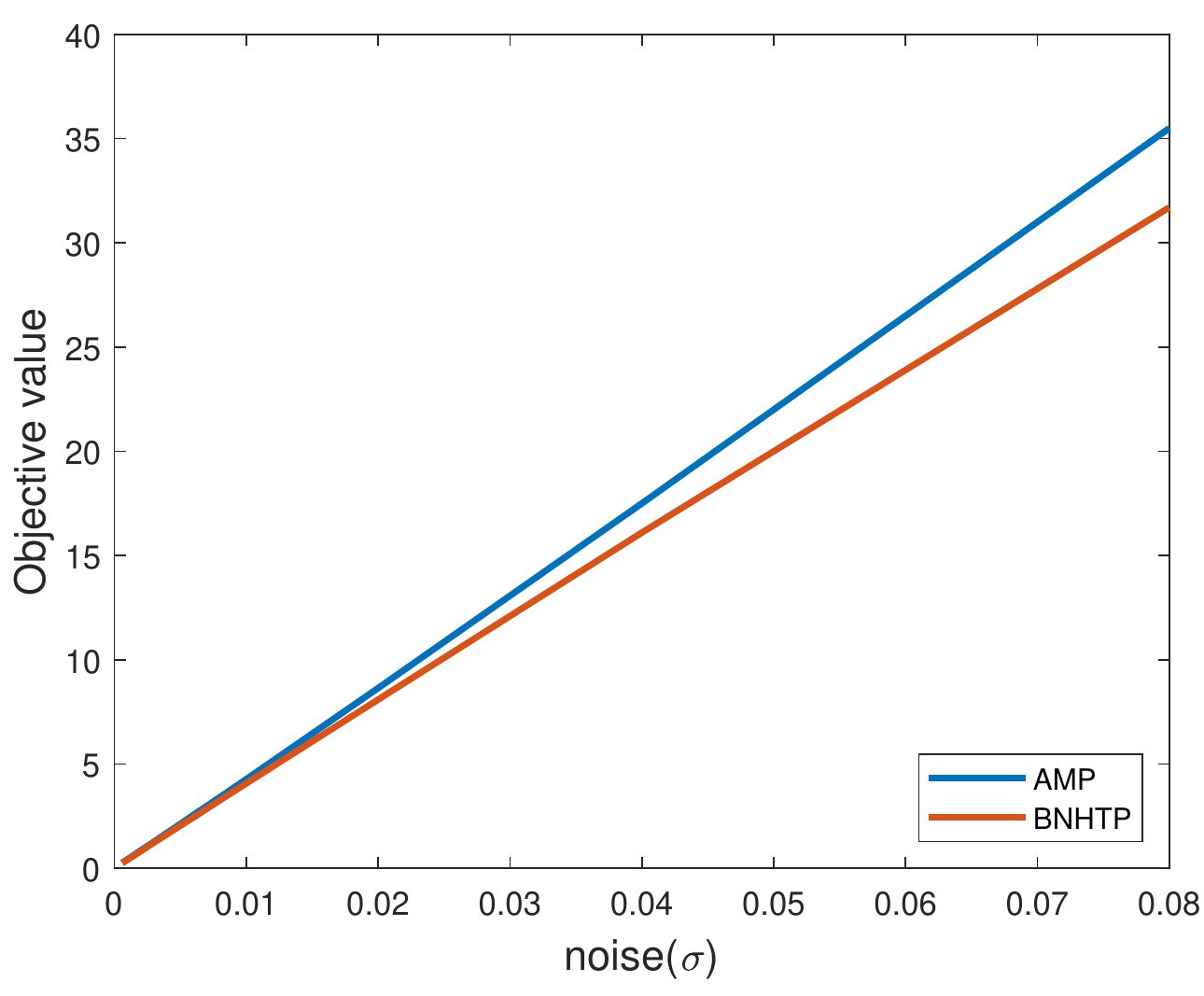}
\includegraphics[height=7cm,width=7cm]{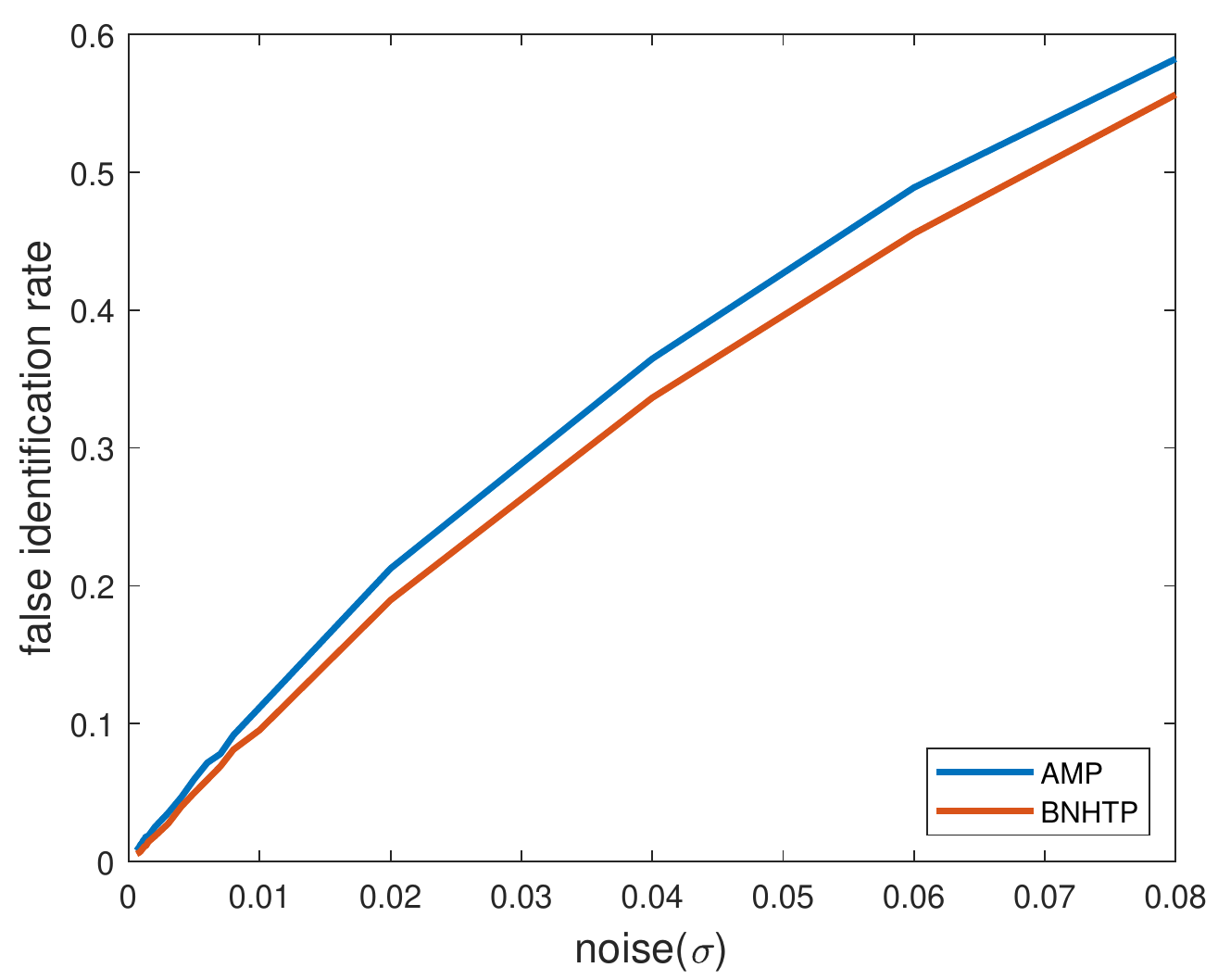}
\caption{Two Algorithms for exponential type (II) matrix with the different noise level.}
\label{figurefinal}
\end{figure}

\begin{table}[H]
\centering
\caption{ Two Algorithms for exponential type (II) matrix with different $\bar s$}
\begin{tabular}{cccccccc}
\hline $\bar s$ & Method &  Iter & Time(${s}$) & R-error &  Obj-value & T rate (\%) & Tc rate(\%) \\
\hline
\multirow{2}{*}{10}
&AMP  & 20.12 & 0.563   & 0.0844   & 0.454   & 97.20 & 99.95 \\
&BNHTP &  {\bf 3.00} &  {\bf 0.049}   &{\bf  0.0361}   & {\bf 0.415}   & {\bf 100.00} & {\bf 100.00} \\
\hline
\multirow{2}{*}{20}
&AMP   &  26.74 & 0.719   & 0.0654   & 0.440   & 94.25 & 99.96 \\
&BNHTP & {\bf 3.00} & {\bf 0.052}   & {\bf 0.0358}   &{\bf  0.411}   &{\bf  100.00} & {\bf 100.00} \\
\hline
\multirow{2}{*}{30}
&AMP   & 27.05 & 0.723   & 0.0942   & 0.545   & 99.97 & 99.87 \\
&BNHTP & {\bf 3.00} &{\bf  0.051}   &{\bf  0.0253}   & {\bf 0.405}   & {\bf 100.00} &{\bf  100.00} \\
\hline
\multirow{2}{*}{40}
&AMP   &  32.85 & 0.873   & 0.0751   & 0.464   & 96.50 & 99.93 \\
&BNHTP &{\bf  3.71} & {\bf 0.067}  & {\bf 0.0366}   & {\bf 0.400}   & {\bf 100.00} & {\bf 100.00} \\
\hline
\multirow{2}{*}{50}
&AMP   & 30.81 & 0.812   & 0.0899   & 0.503   & 94.52 & 99.91 \\
&BNHTP & {\bf 3.94} & {\bf 0.074}   & {\bf 0.0293}   & {\bf 0.393}   & {\bf 100.00} & {\bf 100.00} \\
\hline
\multirow{2}{*}{60}
&AMP   & 29.94 & 0.783   & 0.0907   & 0.696   & 98.68 & 99.74 \\
&BNHTP &{\bf  21.00} &{\bf  0.500}   & {\bf 0.0264}   & {\bf 0.374}   & 98.33 & {\bf 99.98} \\
\hline
\end{tabular}\label{diff-s4}
\end{table}
The other inactive users calculate the false alarm probability in this way and finally average the false alarm probability of all inactive users to get the final false alarm probability. Similarly, in 10000 repetitive simulations, User 1 is active in each simulation (that is, all 32 corresponding elements are zero). But after detection by the algorithm, User 1 corresponding 32 elements are non-zero in 9000 simulation results.  User 1 corresponding 32 elements are zero in other simulation results. The probability of missed detection for this user is 10\%. Other active users also calculate the miss probability in this way.  The final miss probability is obtained by taking the average.\\
\indent In the field of communication, often by controlling the false alarm probability within a certain range. The smaller of corresponding miss probability means the better performance of the algorithm. In the experiment, the element in the recovered signal needs to be set to zero with a threshold value to control the false alarm probability. The sensor matrix selects an exponential type (I) matrix in the next experiment. Assume that 20 users are active, i.e., $\bar s$=20. We carry out 2000 times simulation experiments under different noises. By adjusting the set threshold, the false alarm probability is close to 0.1\%, and the record's missed detection probability and threshold are observed. The numerical results are as figure \ref{figurefinal}.

\indent For each noise, 2000 numerical experiments are carried out. To ensure the final calculated false alarm probability at about 0.1\%, for AMP, we set the small elements in the recovered signal to zero by adjusting the threshold value. It can be found that the probability of miss detection under the BNHTP algorithm is smaller, which is an important problem in the areas of wireless communications. As the noise increases, the BNHTP performs better than the AMP algorithm in terms of the number of iterations, CPU time and relative error.

\section{Conclusion} \label{Section-conclusion}
In this paper, we make progress on compressed sensing with block sparse structure in complex fields. We established a new block sparse constraint optimization model and analysed the properties of the objective function. Furthermore, we defined a new stationary point, analysed first-order sufficient and necessary conditions for the new model and design a new efficient second-order BNHTP algorithm. Numerical experiments demonstrate that the BNHTP algorithm has superior performance in terms of recovery accuracy and calculation time when compared with the AMP algorithm. Does this algorithm have better theoretical properties such as convergence, and can be applied to more complex situations? We leave these interesting questions in our further study.

\section*{Acknowledgements}
This work is supported by the National Natural Science Foundation of China (No.12131004, No.11971052).

\vskip 0.2in
\bibliographystyle{ieeetr} 

\bibliography{refsco}

\setlength{\bibsep}{0pt plus 0.0ex}
\bibliographystyle{elsariticle-num}
\begin{thebibliography}{99}



\bibitem[Andrew(2007)]{Andrew2007}
G. Andrew, J. Gao.
\newblock Scalable training of $l_1$-regularized log-linear models.
\newblock {\em Proceedings of the 24th International Conference on Machine Learning}, 2007:33-40.

\bibitem[Applebaum(2010)]{Applebaum2010}
L. Applebaum, W. U. Bajwa, M. F. Duarte, et al.
\newblock Multiuser detection in asynchronous on-off random access channels using lasso.
\newblock {\em Annual Allerton Conference on Communication, Control and Computing}, 2010:130-137.

\bibitem[Baraniuk(2010)]{Baraniuk2010}
 R. G. Baraniuk, V. Cevher, M. F. Duarte, C. Hegde.
\newblock Model-based compressive sensing,
\newblock {\em IEEE Transactions on Communication}, 56(4):1982-2001, 2010. 

\bibitem[Borgnat(2008)]{Borgnat2008}
 P. Borgnat, P. Flandrin.
\newblock Time-frequency localization from sparsity constraints.
\newblock {\em IEEE International Conference on Acoustics, Speech and Signal Processing}. 2008:3785-3788.

\bibitem[Candes(2006a)]{Cand2006a}
E. J. Candes, T. Tao.
\newblock Compressive sampling.
\newblock {\em Proceedings of the International Congress of Mathematicians}, 3:1433-1452, 2006.

\bibitem[Candes(2006b)]{Cand2006b}
E. J. Candes, J. Romberg, T. Tao.
\newblock Robust uncertainty principles: exact signal reconstruction from highly incomplete frequency information.
\newblock {\em IEEE Transactions on Information Theory},  52(2):489-509, 2006.

\bibitem[Chen(2001)]{Chen2001}
S. Chen, D. L. Donoho, M. A. Saunders.
\newblock Atomic decomposition by basis pursuit.
\newblock {\em Society for Industrial and Applied Mathematics Review}, 43(1):129-159, 2001.

\bibitem[Cotter(2005)]{Cotter2005}
S. F. Cotter, B. D. Rao, K. Engan, et al.
\newblock Sparse solutions to linear inverse problems with multiple measurement vectors.
\newblock {\em IEEE Transactions on Signal Processing}, 53(7): 2477-2488, 2005.

\bibitem[Chen(2006)]{Chen2006}
J. Chen, X. Huo.
\newblock Theoretical results on sparse representations of multiple-measurement vectors.
\newblock {\em IEEE Transactions on Signal Processing}, 54(12): 4634-4643, 2006.

\bibitem[Donoho(2006)]{Donoho2006}
D. L. Donoho.
\newblock Compressed sensing.
\newblock {\em IEEE Transactions on Information Theory}, 52(4):1289-1306, 2006.

\bibitem[Donoho(2009)]{Donoho2009}
D. L. Donoho, A. Maleki, A. Montanari.
\newblock Message-passing algorithms for compressed sensing.
\newblock {\em Proceedings of the National Academy of Sciences}, 106(45):18914-18919, 2009.

\bibitem[Eldar(2009a)]{Eldar2009a}
Y. C. Eldar, M. Mishali.
\newblock Robust recovery of signals from a structured union of subspaces.
\newblock {\em IEEE Transactions on Information Theory}, 55(11): 5302-5316, 2009.

\bibitem[Eldar(2009b)]{Eldar2009b}
Y. C. Eldar, H. Rauhut.
\newblock Average case analysis of multichannel sparse recovery using convex relaxation.
\newblock {\em IEEE Transactions on Information Theory}, 56(1): 505-519, 2009, 

\bibitem[Eldar(2009c)]{Eldar2009c}
 Y. C. Eldar, M. Mishali.
\newblock Block sparsity and sampling over a union of subspaces.
\newblock {\em 2009 16th International Conference on Digital Signal Processing}, 2009: 1-8.

\bibitem[Eldar(2010)]{Eldar2010}
 Y. C. Eldar, P. Kuppinger. 
\newblock Block-sparse signals: uncertainty relations and efficient recovery.
\newblock {\em IEEE Transactions on Signal Processing}, 58(6):3042-3054, 2010.


\bibitem[Hamidi(2010)]{Hamidi2010}
S. Hamidi-Ghalehjegh, M. Babaie-Zadeh, C. Jutten.
\newblock Fast block-sparse decomposition based on SL0.
\newblock {\em Proceedings of 9th International Conference on Latent Variable Analysis and Signal Separation}, 2010: 426-433. 

\bibitem[Huang(2015)]{Huang2015}
 B. Huang, T. Zhou, 
\newblock Recovery of block sparse signals by a block version of StOMP.
\newblock {\em Signal Processing}, 106:231-244, 2015. 

\bibitem[Landau(1967)]{Landau1967}
H. J. Landau.
\newblock Necessary density conditions for sampling and interpolation of certain entire functions.
\newblock {\em Acta Mathematica}, 117:37-52, 1967.

\bibitem[Lustig(2007)]{Lustig2007}
M. Lustig, D. Donoho.
\newblock Sparse MRI: The application of compressed sensing for rapid MR imaging.
\newblock {\em Magnetic Resonance in Medicine: An Official Journal of the International Society for Magnetic Resonance in Medicine}, 58(6):1182-1195, 2007.

\bibitem[Li(2016)]{Li2016}
X. Li, D. Sun,  K. C. Toh.
\newblock A highly efficient semismooth Newton augmented Lagrangian method for solving Lasso problems.
\newblock {\em SIAM Journal on Optimization}, 28(1):433-458, 2018.

\bibitem[Karlina(2011)]{Karlina2011}
R. Karlina, M. Sato.
\newblock Compressive sensing applied to imaging by ground-based polarimetric SAR.
\newblock {\em 2011 IEEE International Geoscience and Remote Sensing Symposium. IEEE}, 2011:2861-2864.

\bibitem[Kutyniok(2013)]{Kutyniok2013}
G. Kutyniok.
\newblock Theory and applications of compressed sensing.
\newblock {\em GAMM-Mitteilungen}, 36(1):79-101, 2013.

\bibitem[Krishnapuram(2005)]{Krishnapuram2005}
B. Krishnapuram,L. Carin, M. A. Figueiredo, et al.
\newblock Sparse multinomial logistic regression: fast algorithms and generalization bounds.
\newblock {\em IEEE Transactions on Pattern Analysis and Machine Intelligence}, 27(6):957-968, 2005. 

\bibitem[Mallat(1993)]{Mallat1993}
S. G. Mallat, Z. Zhang.
\newblock Matching pursuits with time-frequency dictionaries.
\newblock {\em IEEE Transactions on Signal Processing}, 41(12):3397-3415, 1993.

\bibitem[Mishali(2008)]{Mishali2008}
M. Mishali, Y. C. Eldar.
\newblock Reduce and boost: Recovering arbitrary sets of jointly sparse vectors.
\newblock {\em IEEE Transactions on Signal Processing}, 56(10):4692-4702, 2008.

\bibitem[Mishali(2009)]{Mishali2009}
M. Mishali, Y. C. Eldar.
\newblock Blind multiband signal reconstruction: Compressed sensing for analog signals.
\newblock {\em IEEE Transactions on Signal Processing}, 57(3):993-1009, 2009. 

\bibitem[Mishali(2010)]{Mishali2010}
M. Mishali, Y. C. Eldar.
\newblock From theory to practice: Sub-Nyquist sampling of sparse wideband analog signals.
\newblock {\em IEEE Journal of selected topics in signal processing}, 4(2):375-391, 2010.

\bibitem[Mishali(2011)]{Mishali2011}
 M. Mishali, Y. C. Eldar, O. Dounaevsky , et al.
\newblock Xampling: Analog to digital at sub-Nyquist rates.
\newblock {\em IET circuits, devices and systems}, 5(1):8-20, 2011. 

\bibitem[Natarajan(1995)]{Natarajan1995}
 B. K. Natarajan.
\newblock Sparse approximate solutions to linear systems. 
\newblock {\em SIAM Journal on Computing}, 24(2):227–234, 1995.
\bibitem[Parvaresh(2008)]{Parvaresh2008}
F. Parvaresh, H. Vikalo, S. Misra, et al.
\newblock Recovering sparse signals using sparse measurement matrices in compressed DNA microarrays.
\newblock {\em IEEE Journal of Selected Topics in Signal Processing}, 2(3): 275-285, 2008.

\bibitem[Qaisar(2013)]{Qaisar2013}
S. Qaisar, R. M. Bilal, W. Iqbal, et al.
\newblock Compressive sensing: From theory to applications, a survey.
\newblock {\em Journal of Communications and Networks}, 15(5):443-456, 2013.

\bibitem[Sheikh(2007)]{Sheikh2007}
M. A. Sheikh, S. Sarvotham, O. Milenkovic, et al.
\newblock DNA array decoding from nonlinear measurements by belief propagation.
\newblock {\em IEEE 14th Workshop on Statistical Signal Processing}. IEEE, 2007:215-219.

\bibitem[Tibshirani(1996)]{Tibshirani1996}
R. J. Tibshirani.
\newblock Regression shrinkage and selection via the LASSO.
\newblock {\em Journal of the Royal Statistical Society. Series B: Methodological}, 73(1):273-282, 1996.

\bibitem[Wang(2013)]{Wang2013}
Y. Wang, J. Wang, Z. Xu.
\newblock On recovery of block-sparse signals via mixed $l_2/l_q(0<q\leq$1) norm
minimization,
\newblock {\em EURASIP Journal on Advances in Signal Processing}, 76:1-17, 2013. 

\bibitem[Xie(2013)]{Xie2013}
Y. Xie, Y. C. Eldar, A. Goldsmith.
\newblock Reduced-dimension multiuser detection.
\newblock {\em IEEE Transactions on Information Theory}, 59(6):3858-3874, 2013.

\bibitem[Xu(2018)]{Xu2018}
G. Xu, Y. Liu, M. Xing.
\newblock Multi-channel synthetic aperture radar imaging of ground moving targets using compressive sensing.
\newblock {\em IEEE Access}, 6:66134-66142, 2018.

\bibitem[Yin(2013)]{Yin2013}
H. Yin, S. Li, L. Fang.
\newblock Block-sparse compressed sensing: non-convex model and iterative reweighted algorithm.
\newblock {\em Inverse Problems in Science and Engineering}, 21(1):141-154, 2013.

\bibitem[Zeinalkhani(2015)]{Zeinalkhani2015}
Z. Zeinalkhani, A. Banihashemi.
\newblock Iterative reweighted $l_2/l_1$ recovery algorithms for compressed sensing 
of block sparse signals.
\newblock {\em IEEE Transactions on Signal Processing}, 63(17): 4516-4531, 2015. 

\bibitem[Zhang(2014)]{Zhang2014}
X. Zhang, M. Li, L. Zuo.
\newblock Compressed sensing detector for wideband radar using the dominant scatterer.
\newblock {\em IEEE Signal Processing Letters}, 21(10):1275-1279, 2014.

\bibitem[Zhou(2021)]{Zhou2021}
S. Zhou, L. Pan, N. Xiu.
\newblock Newton method for $l_0$-regularized optimization.
\newblock {\em Numerical Algorithms}, 88(4):1541-1570, 2021.

\bibitem[Zhou(2019)]{Zhou2019}
S. Zhou, N. Xiu, H. Qi.
\newblock Global and quadratic convergence of Newton hard-thresholding pursuit.
\newblock {\em Journal of Machine Learning Research}, 22(12):1-45, 2019.

\end{thebibliography}

\section*{APPENDIX: PROOFS OF THEORETICAL RESULTS}
In this appendix, we provide complete proof of theoretical results.\\
{\bf A. Proof of Lemma \ref{f-Lips}}\\
{\bf Proof:} From the mean value theorem, for $\forall \p, \q \in \Cn$, there exists vector $\varphi=\p+t(\q-\p), t \in(0,1)$, such that
\[\nabla f(\p)-\nabla f(\q)=\nabla^{2} f(\varphi)(\p-\q).\]
Together with
\[\|\nabla f(\varphi)\| \leq \sqrt{2 \lambda_{\max }(\A^{\rm{H}} \A)},\]
we can deduce
\begin{align*}
\left\|\nabla f(\p)-\nabla f(\q)\right\| & \leq\left\|\nabla^{2} f(\varphi)\right\|\left\|\p-\q\right\|  \leq 2 \lambda_{\max }(\A^{\rm{H}} \A)\left\|\p-\q\right\|=\frac{1}{\alpha_{f}}\left\|\p-\q\right\|.
\end{align*}
The result is on hand.\hfill $\square$\\
\noindent {\bf B. Proof of Lemma \ref{f-descent}.}\\
 {\bf Proof:} Let $g(t)=f(\q+t(\p-\q))$, $t\in(0,1)$, then we have $\dfrac{dg(t)}{dt}=\dfrac{df(\q+t(\p-\q))}{dt}$. Pick a vector $u=\q+t(\p-\q)$.  Then we can easily get
\[df(u)=\frac{\partial f(u)}{\partial u}  d u+\frac{\partial f(u)}{\partial \bar{u}}  d \bar{u}\]
Then we have
\begin{align*}
\dfrac{dg(t)}{dt}&=\frac{\partial f(u)}{\partial u}\dfrac{ d u}{dt}+\frac{\partial f(u)}{\partial \bar{u}} \dfrac{d \bar{u}}{dt}\\
& =[(\p-\q)^{\rm{H}} \frac{\partial f(\q+t(\p-\q))}{\partial \q}+(\overline{\p}-\overline{\q})^{\rm{H}} \frac{\partial f(\q+t(\p-\q))}{\partial \overline{\q}}]\dfrac{1}{dt}.
\end{align*}
So we can get
\begin{align*}
f(\p)-f(\q)&=g(1)-g(0)\\
&=\int_{0}^{1}\left[(\p-\q)^{\rm{H}} \frac{\partial f(\q+t(\p-\q))}{\partial \q}+(\overline{\p}-\overline{\q})^{\rm{H}} \frac{\partial f(\q+t(\p-\q))}{\partial \overline{\q}}\right] d t .
\end{align*}
Therefore, it can be obtained
\begin{align*}
f(\p)=f(\q)+\int_{0}^{1}\left[(\p-\q)^{\rm{H}} \nabla f(\q+t(\p-\q))
+(\overline{\p}-\overline{\q})^{\rm{H}} \overline{\nabla f(\q+t(\p-\q))}\right] d t.
\end{align*}
Further, it can be written as
\begin{align*}
f(\p)  =& f(\q)+(\p-\q)^{\rm{H}} \nabla f(\q)+(\overline{\p}-\overline{\q})^{\rm{H}} \overline{\nabla f(\q)} +\int_{0}^{1}\left[(\p-\q)^{\rm{H}}(\nabla f(\q+t(\p-\q))-\nabla f(\q))\right.\\
&\left.+(\overline{\p}-\overline{\q})^{\rm{H}}(\overline{\nabla f(\q+t(\p-\q))}-\overline{\nabla f(\q)})\right] d t.
\end{align*}
Taking the norm on both sides and combining it with the Lemma \ref{f-Lips}, we get
\begin{align*}
&f(\p)-f(\q)-(\p-\q)^{\rm{H}} f(\q)-(\overline{\p}-\overline{\q})^{\rm{H}} \overline{\nabla f(\q)} \\
=&\int_{0}^{1} ( (\p-\q)^{\rm{H}} (\nabla f(\q+t(\p-\q)) - \nabla f(\q)) \\
&\left.\quad+(\overline{\p}-\overline{\q})^{\rm{H}}(\overline{\nabla f(\q+t(\p-\q))}-\overline{\nabla f(\q)})\right] d t \\
\leq &\left\|\int_{0}^{1}(\p-\q)^{\rm{H}}(\nabla f(\q+t(\p-\q))-\nabla f(\q)) d t\right\| \\
& \quad+\left\|\int_{0}^{1}(\overline{\p}-\overline{\q})^{\rm{H}}(\overline{\nabla f(\q+t(\p-\q))}-\overline{\nabla f(\q)}) d t\right\| \\
\leq & \int_{0}^{1}\left\|\p-\q\right\|\left\|\nabla f(\q+t(\p-\q))-\nabla f(\q)\right\| d t \\
& \quad+\int_{0}^{1}\left\|\overline{\p}-\overline{\q}\right\|\left\|\overline{\nabla f(\q+t(\p-\q))}-\overline{\nabla f(\q)}\right\| d t \\
\leq & \int_{0}^{1} \frac{t}{\alpha_f}\left\|\p-\q\right\|^{2} d t+\int_{0}^{1} \frac{t}{\alpha_f}\left\|\overline{\p}-\overline{\q}\right\|^{2} d t \\
=& \frac{1}{2 \alpha_f}\left\|\p-\q\right\|^{2}+\frac{1}{2 \alpha_f}\left\|\overline{\p}-\overline{\q}\right\|^{2} .
\end{align*}
Due to $\left\|\p-\q\right\|^{2}=\left\|\bar{\p}-\bar{\q}\right\|^{2}$, we have
\begin{align*}
f(\p)-f(\q)-(\p-\q)^{\rm{H}} \nabla f(\q)-(\overline{\p}-\overline{\q})^{\rm{H}} \overline{\nabla f(\q)} \leq \frac{1}{\alpha_f}\left\|\p-\q\right\|^{2} .
\end{align*}
Therefore, we get the conclusion.  \hfill $\square$\\
\noindent{\bf C. Proof of Lemma \ref{lemma3}.}\\
Proof: From the proof process of lemma \ref{f-descent}, there exist a $t\in(0,1)$ such that
\begin{align*}
f(\p)=f(\q)+\int_{0}^{1}\left[(\p-\q)^{\rm{H}} \nabla f(\q+t(\p-\q))
+(\overline{\p}-\overline{\q})^{\rm{H}} \overline{\nabla f(\q+t(\p-\q))}\right] dt, \forall \p, \q\in \Cn.
\end{align*}
Hence, we just need to prove
\[\begin{aligned}
\int_{0}^{1}\left[(\p-\q)^{\rm{H}}(\nabla f(\q+t(\p-\q))-\nabla f(\q))\right.
\left.+(\overline{\p}-\overline{\q})^{\rm{H}}(\overline{\nabla f(\q+t(\p-\q))}-\overline{\nabla f(\q)})\right] d t\geq 0 .
\end{aligned}\]
According to the mean value theorem, there exists a $\lambda(t) \in(0,1)$, $\varphi (t)=\lambda(t) \q+(1-\lambda(t) )(\q+t(\p-\q))=\q+\lambda(t)  t(\p-\q), \forall t \in(0,1)$, which makes
\[\nabla f(\q+t(\p-\q))-\nabla f(\q)
=\lambda(t)t\nabla^{2} f(\varphi (t))(\p-\q),\]
\[\overline{\nabla f(\q+t(\p-\q))}-\overline{\nabla f(\q)}
=\lambda(t)t\overline{\nabla^{2} f(\varphi (t))}(\overline{\p}-\overline{\q}).\]
Since $\lambda(t)>0$, $t>0$, $\nabla^{2} f(\varphi (t))= \A^{H}\A$, $\overline{\nabla^{2} f(\varphi (t))}=\overline{ \A}^{H}\overline{ \A}$,
We have
\[\begin{aligned}
& \quad \int_{0}^{1}\left[(\p-\q)^{\rm{H}}(\nabla f(\q+t(\p-\q))-\nabla f(\q))\right.
\left.+(\overline{\p}-\overline{\q})^{\rm{H}}(\overline{\nabla f(\q+t(\p-\q))}-\overline{\nabla f(\q)})\right] d t\\
& =\int_{0}^{1}\left[(\p-\q)^{\rm{H}}(\lambda(t)t\nabla^{2} f(\varphi (t))(\p-\q))\right.
\left.+(\overline{\p}-\overline{\q})^{\rm{H}}(\lambda(t)t\overline{\nabla^{2} f(\varphi (t))}(\overline{\p}-\overline{\q}))\right] d t\\
& = \int_{0}^{1}\left[\lambda(t)t\left[(\p-\q)^{\rm{H}}(\nabla^{2} f(\varphi (t))(\p-\q))\right.
\left.+(\overline{\p}-\overline{\q})^{\rm{H}}(\overline{\nabla^{2} f(\varphi (t))}(\overline{\p}-\overline{\q}))\right]\right]d t.\\
& = \int_{0}^{1}\left[\lambda(t)t\left[(\p-\q)^{\rm{H}}(\A^{H}\A)(\p-\q)\right.
\left.+(\overline{\p}-\overline{\q})^{\rm{H}}(\overline{ \A}^{H}\overline{ \A})(\overline{\p}-\overline{\q})\right]\right]d t,
\end{aligned}\]
and
\[\begin{aligned}
&(\p-\q)^{\rm{H}}(\A^{H}\A)(\p-\q)+(\overline{\p}-\overline{\q})^{\rm{H}}(\overline{ \A}^{H}\overline{\A})(\overline{\p}-\overline{\q})\\
=&2\langle\A(\p-\q),~\A(\p-\q) \rangle=2\|\A(\p-\q)\|^2 \geq 0.
\end{aligned}\]
Then it is can be seen that
\[\int_{0}^{1}\left[(\p-\q)^{\rm{H}}(\nabla f(\q+t(\p-\q))-\nabla f(\q))\right.
\left.+(\overline{\p}-\overline{\q})^{\rm{H}}(\overline{\nabla f(\q+t(\p-\q))}-\overline{\nabla f(\q)})\right] d t\geq 0.\]
So we can get
\[f(\p)  \geq f(\q)+(\p-\q)^{\rm{H}} \nabla f(\q)+(\overline{\p}-\overline{\q})^{\rm{H}} \overline{\nabla f(\q)}.\]
Hence we get the conclusion.  \hfill $\square$\\

\noindent{\bf D. Proof of Theorem \ref{first-order}}\\
Proof: ($\Rightarrow$) Suppose that $\x^{*}\in\Cn$ is the optimal solution of the model (\ref{BCS}), and for $i\in \widehat{T}\subset\{1,2,\ldots ,I\}$, $ \widehat{T}\neq\emptyset$, there exists a $\x^{*}_{[i]} \notin {\rm P}_{\S_{[i]}}((\x^{*}-\tau\nabla f(\x^{*}))_{[i]})$ which means that $\x^{*}$ is not a $\tau$-stationary point of the model (\ref{BCS}). So we just need to deduce that $\x^{*}$ is not the optimal solution of the model (\ref{BCS}), which is contradictory to the assumption.\\
\indent Let $\x\in\Cn$ is $\tau$-stationary point of the model (\ref{BCS}). According to the nature of projection, for $i\in \widehat{T}$, we have
\begin{align*}
\left\|\x_{[i]}-(\x^{*}-\tau \nabla f(\x^{*}))_{[i]}\right\|^{2} & < \left\|\x_{[i]}^{*}-(\x^{*}-\tau \nabla f(\x^{*}))_{[i]}\right\|^{2}.
\end{align*}
This leads to the following expression: 
\begin{align*}
\left\|\x_{[i]}-\x^{*}_{[i]}\right\|^{2}+\left\|\tau\nabla f(\x^{*}))_{[i]}\right\|^{2}+2 \langle(\x_{[i]}-\x^{*}_{[i]}),~\tau\nabla f(\x^{*}))_{[i]}\rangle & < \left\|\tau f(\x^{*}))_{[i]}\right\|^{2}, \\
\end{align*}
and
\begin{align*}
&\langle (\x-\x^{*})_{[i]},~ \nabla f(\x^{*})_{[i]} \rangle<-\frac{1}{2 \tau}\left\|(\x-\x^{*})_{[i]}\right\|^{2}.
\end{align*}
Based on the fact
\begin{align*}
\sum\limits_{i=1}^{\boldsymbol{I}}\langle (\x-\x^{*})_{[i]},~ \nabla f(\x^{*})_{[i]} \rangle=\langle (\x-\x^{*}),~ \nabla f(\x^{*}) \rangle,
\end{align*}
$$\langle (\x-\x^{*}),~ \nabla f(\x^{*}) \rangle =\frac{1}{2}(\x-\x^{*})^{\rm{H}} \nabla f(\x^{*})+\frac{1}{2}(\overline{\x}-\overline{\x}^{*})^{\rm{H}} \overline{\nabla f(\x^{*})},$$
$$\sum\limits_{i=1}^{\boldsymbol{I}}\left\|(\x-\x^{*})_{[i]}\right\|^{2}= \left\|(\x-\x^{*})\right\|^{2}, $$
and according to lemma \ref{f-descent} and $\tau \leq\alpha_{f}$,  we  obtain
\begin{align*}
f(\x) & \leq f(\x^{*})+(\x-\x^{*})^{\rm{H}} \nabla f(\x^{*})+(\overline{\x}-\overline{\x}^{*})^{\rm{H}} \overline{\nabla f(\x^{*})}+\frac{1}{\alpha_{f}}\left\|\x-\x^{*}\right\|^{2} \\
& < f(\x^{*})+(\frac{1}{\alpha_{f}}-\frac{1}{\tau})\left\|\x-\x^{*}\right\|^{2} \\
&<f(\x^{*}),
\end{align*}
which means $\x^{*}$ is not an optimal solution for the model (\ref{BCS}). Then based on proof by contradiction, the conclusion is on hand.\\
($\Leftarrow$) We divide this proof into two parts as follows:\\
 {\bf Case(i)} $\forall i$, $\left\|{\x_{[i]}^{*}}\right\|_{0}=s_i$. In this case, for any $\x\in \C_{\widehat{\Gamma}}^{n}$, we can see that $\x_{j}^{*}=\x_{j}=0$, where $j \notin \widehat{\Gamma}_{[i]}$. And $(\nabla f(\x^{*}))_{[i]_{j}}=0$ if $j\in \widehat{\Gamma}_{[i]}$, we have $(\overline{\nabla f(\x^{*})})_{[i]_{j}}=0$.
It follows from Lemma \ref{lemma3} that
\begin{align*}
f(\x) \geq& f(\x^{*})+(\x-\x^{*})^{\rm{H}} \nabla f(\x^{*})+ (\bar{\x}-\bar{\x}^{*})^{\rm{H}} \overline{\nabla f(\x^{*})} \\
=& f(\x^{*})+\sum_{i =1} (\x_{[i]}-\x_{[i]}^{*})^{\rm{H}}(\nabla f(\x^{*}))_{[i]}+ \sum_{i =1} (\bar{\x}_{[i]}-\bar{\x}_{[i]}^{*})^{\rm{H}}(\overline{\nabla f(\x^{*})} )_{[i]} \\
=& f(\x^{*})+ \sum_{i =1} (\sum_{j \in\widehat{\Gamma}_{[i]}} ( \x_{[i]_{j}}-\x_{[i]_{j}}^{*})^{\rm{H}}\nabla f(\x^{*})_{[i]_{j}}+\sum_{j \in\widehat{\Gamma}_{[i]}} ( \x_{[i]_{j}}-\x_{[i]_{j}}^{*})^{\rm{H}}\nabla f(\x^{*})_{[i]_{j}})\\
&+ \sum_{i =1} (\sum_{j \in\widehat{\Gamma}_{[i]}} ( \bar{\x}_{[i]_{j}}-\bar{\x}_{[i]_{j}}^{*})^{\rm{H}}\overline{\nabla f(\x^{*})_{[i]_{j}}}+\sum_{j \in\widehat{\Gamma}_{[i]}} ( \bar{\x}_{[i]_{j}}-\bar{\x}_{[i]_{j}}^{*})^{\rm{H}}\overline{\nabla f(\x^{*})_{[i]_{j}}})\\
=& f(\x^{*}) .
\end{align*}
So $\x^{*}$ is the optimal solution of the the model (\ref{BCS}) in subspace $\C_{\widehat{\Gamma}}^{n}$ .

{\bf Case(ii)} $\|{\x_{[i]}^{*}}\|_{0}<s_i$ for all $i\in\{1,2,...,I\}$. We have $\nabla f(\x^{*})_{[i]}=0$ in this case. Similar to the proof process of {\bf case (i)}, we have
\begin{align*}
f(\x) \geq f(\x^{*})+ (\x-\x^{*})^{\rm{H}} \nabla f(\x^{*})+ (\bar{\x}-\bar{\x}^{*})^{\rm{H}} \overline{\nabla f(\x^{*})} =f(\x^{*}) , \forall \x \in \boldsymbol{S},
\end{align*}
which also displays $\x^{*} $ is the optimal solution of the model (\ref{BCS}). \hfill $\square$\\

\noindent{\bf E. Four methods to generate $\A$ mentioned in section \ref{Section-Numerical}} \\
\noindent (i) Gaussian complex matrix $\A$. Let $\A \in \C^{m \times n}$  be a random gaussian complex matrix. Each column $\a_{j} \in \C^{m \times 1} \sim \mathcal{C} \mathcal{N}(\mathbf{0}, \boldsymbol{I}), j \in {1,2,...,n}$ satisfies independent of the same standard normal distribution.\\
\noindent (ii) Partial discrete cosine transform complex matrix.  Let  $\A\in\C^{839\times 2048}$ be a random partial discrete cosine transform complex matrix generated by
\[
\A_{mn}=\cos (2 \pi n-1) \psi_{m})+ \rm{i} * \cos (2 \pi(n-1) \psi_{n}), \quad m=1, \ldots, 839, \quad n=1, \ldots, 2048,
\]
where $\psi_{i}$ and $\psi_{k}$ are uniformly and independently sampled from [0,1].\\
\noindent (iii) Exponential type (I) complex matrix $\A\in\C^{839\times2048}$:
\begin{align*}
 \A_{mn} = \begin{cases}
e^{j \frac{420 \pi m(m-1)}{839}} e^{j \frac{2 \pi(m-1)(n-1)}{839}},      & if 1 \leq m \leq 839,1 \leq n \leq 832; \\ 
e^{j \frac{419 \pi m(m-1)}{839}} e^{j \frac{2 \pi(m-1)(n-833)}{839}}, & if 1 \leq m \leq 839,833 \leq n \leq 1664;\\
e^{j \frac{\pi m(m-1)}{839}} e^{j \frac{2 \pi(m-1)(n-1665)}{839}},       & if 1 \leq m \leq 839,1665 \leq n \leq 2048.\end{cases}
\end{align*}
\noindent (iv)  Exponential type (II) complex matrix $\A\in\C^{839\times 5952}$:
\begin{align*}
 \A_{mn} = \begin{cases}
e^{j \frac{420 \pi m(m-1)}{839}} e^{j \frac{2 \pi(m-1)(n-1)}{839}},        & if 1 \leq m \leq 839,1 \leq n \leq 837; \\ 
e^{j \frac{419 \pi m(m-1)}{839}} e^{j \frac{2 \pi(m-1)(n-838)}{839}},   & if 1 \leq m \leq 839,838 \leq n \leq 1674;\\
e^{j \frac{\pi m(m-1)}{839}} e^{j \frac{2 \pi(m-1)(n-1675)}{839}},         & if 1 \leq m \leq 839,1675 \leq n \leq 2511;\\
e^{j \frac{838 \pi m(m-1)}{839}} e^{j \frac{2 \pi(m-1)(n-2512)}{839}}, & if 1 \leq m \leq 839,2512 \leq n \leq 3348;\\
e^{j \frac{15 \pi m(m-1)}{839}} e^{j \frac{2 \pi(m-1)(n-3349)}{839}},   & if 1 \leq m \leq 839,3349 \leq n \leq 4185;\\
e^{j \frac{824 \pi m(m-1)}{839}} e^{j \frac{2 \pi(m-1)(n-4186)}{839}}, &  if  1 \leq m \leq 839,4186 \leq n \leq 5022;\\
e^{j \frac{427 \pi m(m-1)}{839}} e^{j \frac{2 \pi(m-1)(n-5023)}{839}},  & if  1 \leq m \leq 839,5023 \leq n \leq 5859;\\
e^{j \frac{412 \pi m(m-1)}{839}} e^{j \frac{2 \pi(m-1)(n-5860)}{839}},  & if 1 \leq m \leq 839,5860 \leq n \leq 5952.
\end{cases}
\end{align*}

\end{document}